\documentclass[12pt]{article}
\usepackage{graphicx}

\begin{document}

\title{On comparison of the Earth orientation parameters obtained from different VLBI networks and observing programs}
\author{Zinovy Malkin \\
  Central Astronomical Observatory at Pulkovo RAS, \\ Pulkovskoe~Ch. 65, St.~Petersburg, 196140 Russia \\ malkin@gao.spb.ru}
\date{28 July 2008}
\maketitle

\begin{abstract}
In this paper, a new geometry index of Very Long Baseline Interferometry (VLBI)
observing networks, the volume of network $V$,
is examined as an indicator of the errors in the Earth orientation
parameters (EOP) obtained from VLBI observations. It has been shown that
both EOP precision and accuracy can be well described by the power law $\sigma=aV^c$
in a wide range of the network size from domestic to global VLBI networks.
In other words, as the network volume grows, the EOP errors become smaller
following a power law.
This should be taken into account for a proper comparison of EOP estimates
obtained from different VLBI networks.
Thus performing correct EOP comparison allows us to accurately investigate
finer factors affecting the EOP errors.
In particular, it was found that the dependence of the EOP precision and
accuracy on the recording data rate can also be described by a power law.
One important conclusion is that the EOP accuracy depends primarily on the network
geometry and to lesser extent on other factors, such as recording mode and data
rate and scheduling parameters, whereas these factors have stronger impact
on the EOP precision.
\end{abstract}


\section{Introduction}
\label{intro}

Very Long Baseline Interferometry (VLBI) is a unique technique to determine
the Earth orientation parameters (EOP), because it is the only method, which
can provide the complete set of five EOP: terrestrial pole coordinates $X_p$ and $Y_p$,
Universal Time $UT1$, and celestial pole coordinates $X_c$ and $Y_c$, with highest
accuracy. Therefore, investigation of various influences on the errors in EOP
derived from the VLBI observations is important for improving the results.
One of the main factors affecting the EOP results is the network geometry.

Evaluation of the dependence of the EOP errors on the network geometry
is important for many practical tasks, such as comparison of EOP estimates
obtained on different VLBI networks, observation scheduling, design of VLBI
experiments and networks, optimal use of the existing VLBI stations for
determination of EOP.
For this reason,
an analysis of the differences in EOP values and errors obtained from
different observing programs and networks
is one of the most actual topics in geodetic VLBI, and many
authors have contributed to these studies, e.g. MacMillan and Ma (2000);
Sokolskaya and Skurikhina (2000); Johnson (2004); Lambert and Gontier (2006); Searle (2006).
However the authors of these papers did not investigate directly the immediate
connection between EOP and the quantitative parameters of the network geometry.

As follows from the basic principles of VLBI measurements,
errors in geodetic and astronomical parameters derived from the
VLBI observations directly depend on the number and the length
of the baselines involved.
Let us have a look at the basic VLBI equation, keeping the main terms in the geometric delay:
\begin{equation}
c\tau = \vec{r}\cdot (\vec{Q}\,\vec{e_s}) = r\,(\vec{e_r} \cdot (\vec{Q}\,\vec{e_s}))\,,
\label{eq:delay}
\end{equation}
where $\vec{r}=r\vec{e_r}$ is the baseline vector in terrestrial coordinate system,
$r$ and $\vec{e_r}$  are its length, and corresponding unit vector,
$\vec{e_s}$ is a unit geocentric vector in celestial coordinate system in the direction of the radio source,
$\vec{Q}$ is the rotation matrix for the transformation from celestial to terrestrial coordinate system.
To get a corresponding equation of condition to be used in a parameter solution,
one has to compute the partial derivatives of Eq.~(\ref{eq:delay}) with respect
to the parameters to be solved.
In case of EOP, the partial derivatives are given as
\begin{equation}
\frac{\partial\tau}{\partial p} =
 c^{-1} \, r\, \left( \vec{e_r} \cdot \left( \frac{\partial \vec{Q}}{\partial p} \, \vec{e_s} \right) \right) \,,
\label{eq:partial}
\end{equation}
where $p$ is one of the Earth rotation parameters.

As one can see from Eq.~(\ref{eq:partial}), the partial derivatives of the
interferometric delay with respect to EOP are proportional to the baseline length.
This means that a longer baseline yields smaller EOP errors.
It should be mentioned, however, that since we consider the VLBI observations
carried out on the real near-spherical Earth, the longest baselines compatible with the Earth's diameter
in practice do not necessarily have smallest EOP error due to the problem of the common visibility
of radio sources from the two ends of the baseline. For this reason, an optimal baseline length
can be found for every specific geodetic or astronomical task, but this analysis is out
of the scope of this paper.

To perform a more detailed analysis of the dependence of the EOP errors on the
baseline orientation, one can rewrite Eq.~(\ref{eq:delay}) as
\begin{equation}
c\tau = r_e \cos \delta \cos h + r_p \sin \delta \,,
\label{eq:delay2}
\end{equation}
where $r_e$ and $r_p$ are the equatorial and polar components (projections)
of the baseline vector $\vec{r}$ respectively,
$\delta, h$ being the source declination and hour angle respectively.
Detailed consideration of this equation which can be found e.g.
in Dermanis and Mueller (1978), Ma (1978), Nothnagel et al. (1994), Schuh (2000),
along with a joint analysis of the set of Eqs.~(\ref{eq:delay2}) written for all
baselines of the given station network,
allows us to conclude that larger baseline equatorial component yields
smaller $UT1$ error, and larger both polar and equatorial projections are
needed to get a better separation and less errors in $X_p$, $Y_p$, $X_c$, and $Y_c$.
Also important is that errors in the pole coordinates $X_p$ and $Y_p$ may depend
on the network longitudinal orientation in case of a moderate longitude span
of the network.

In general, one can expect that the sought-for dependence of the EOP errors
on the network geometry is a function of several indices, such as:
\begin{itemize}
\item the network span in various directions e.g. geodetic (latitude $\Delta\varphi$,
   $\Delta\lambda$) or Cartesian ($\Delta X, \Delta Y, \Delta Z$) coordinates;
\item the network orientation, e.g. central longitude, prevailing baselines
   directions;
\item the number of stations or baselines.
\end{itemize}
All these factors definitely influence, to a greater or lesser extent,
the EOP results obtained from VLBI observations.
Hence they should be accounted for in order to provide more rigorous comparison
of the EOP derived from different networks and separate the impact
of the network geometry and other factors, such as recording data rate and
scheduling parameters, on the EOP errors.
However, a large number of the network parameters makes it difficult
to develop a practical method of comparison of EOP.
Therefore, it would be useful and convenient to have a generalized
index of the network geometry which could serve as an argument of
a simple enough but sufficiently accurate function describing the
dependence of the EOP errors on the network geometry, and could
be used in the comparative studies.

In this paper, continuing the previous study (Malkin 2007),
such a generalized index of the VLBI network geometry is considered,
namely the volume of network, more precisely the volume of the polyhedron
with the network stations at the vertices.
In section~\ref{net_vol}, the dependence of the EOP errors on the network volume
has been investigated, and in section~\ref{prog_comp}, this dependence will be
applied to a more rigorous comparison of the EOP errors obtained from different
VLBI networks.

It is important to distinguish between two kind of errors---precision, which is
a measure of the repeatability (reproducibility) of estimates, and accuracy,
which refers to the agreement between our estimates and the true value (veracity).
Hereafter, it is considered that the EOP precision can be characterized by
the formal error (uncertainty) of the EOP estimates computed making
use of a data processing software.
To assess the EOP accuracy one can compare the EOP obtained from
VLBI observations with an independent EOP series provided
by another space geodesy technique.
In this paper, like in other similar studies (e.g. Vennebusch et al. 2007),
we used for comparison the EOP series provided by
the International GNSS Service (IGS, Dow et al. 2005).
The term 'error' is used here as a collective name for precision and accuracy
in the case when both precision and accuracy are considered.

For completeness, mention may be made of another measure of the EOP quality,
the Allan deviation, which may be used either in its original formulation (Gambis 2002)
or with extensions developed in Malkin (2008) for analyzing unequally weighted
and multidimensional observations, e.g. both pole coordinates simultaneously.
This method was not used in this work, although it may be worth trying for
supplement analysis.

For our computations, the VLBI observations were used, collected on the
global and regional networks in the framework of observing
programs coordinated by the International VLBI Service for Geodesy and Astrometry
(IVS, Schl\"uter and Behrend 2007), and stored in the IVS
Data Centers\footnote{http://ivscc.gsfc.nasa.gov/products-data/data.html}.


\section{Dependence of the EOP errors on the network volume}
\label{net_vol}

This study begins with the computation of an about 10.7-year EOP series for
the period from July 1996 till February 2007, 1440 24-hour sessions in total.
The start epoch of the investigated time series was chosen to be the same as
the start epoch of the IGS EOP series igs00p03.erp used hereafter
for accuracy assessment of the VLBI terrestrial pole coordinates.

During the session processing the volume of the session network was computed
simultaneously with EOP in the following way.
\begin{enumerate}
\item Compute the tetrahedron mesh for the network polyhedron
   by means of the Delaunay triangulation making use of the GEOMPACK package by
   B.~Joe (Joe, 1991).
\item Compute the volume of each tetrahedron as a scalar triple product:
\begin{equation}
|(\vec{r_2}-\vec{r_1})\cdot((\vec{r_3}-\vec{r_1})\times(\vec{r_4}-\vec{r_1}))| \ / \ 6 \,,
\label{eq:volume}
\end{equation}
where $\vec{r_1}, \vec{r_2}, \vec{r_3}, \vec{r_4}$ are the geocentric station vectors.

\item Compute the total network volume as the sum of the volumes of all the tetrahedrons.
\end{enumerate}

Some examples of the volume of different IVS observing networks processed
in this work are presented in Table~\ref{tab:volume},
where the smallest and the largest IVS networks are shown.
The first column of the table contains the IVS session name.
The smallest IVS networks are typical of Japanese (JADE) and
European (EUROPE) regional sessions.
The largest networks were observed in T2 experiments
primarily aimed at the improvement of the terrestrial reference system.
One can see from Table~\ref{tab:volume} that the volume of IVS networks
differ from each other by several orders of magnitude.
For comparison, the volume of the Earth is 1083~Mm$^3$.

\begin{table}
\centering
\caption{The volume of the smallest and largest IVS networks}
\label{tab:volume}
\begin{tabular}{lc}
\hline
Session & Volume, Mm$^3$ \\
\hline
JADE-0610 & 9.925E-04 \\
EUROPE-35 & 5.285E-03 \\
JADE-0601 & 1.008E-02 \\
EUROPE-56 & 1.761E-02 \\
\multicolumn{2}{c}{$\dots$} \\
T2037  & 4.607E+02 \\
T2038  & 4.723E+02 \\
T2041  & 4.816E+02 \\
T2043  & 4.871E+02 \\
\hline
\end{tabular}
\end{table}

\begin{figure}
\centering
\resizebox{0.7\hsize}{!}{\includegraphics[clip]{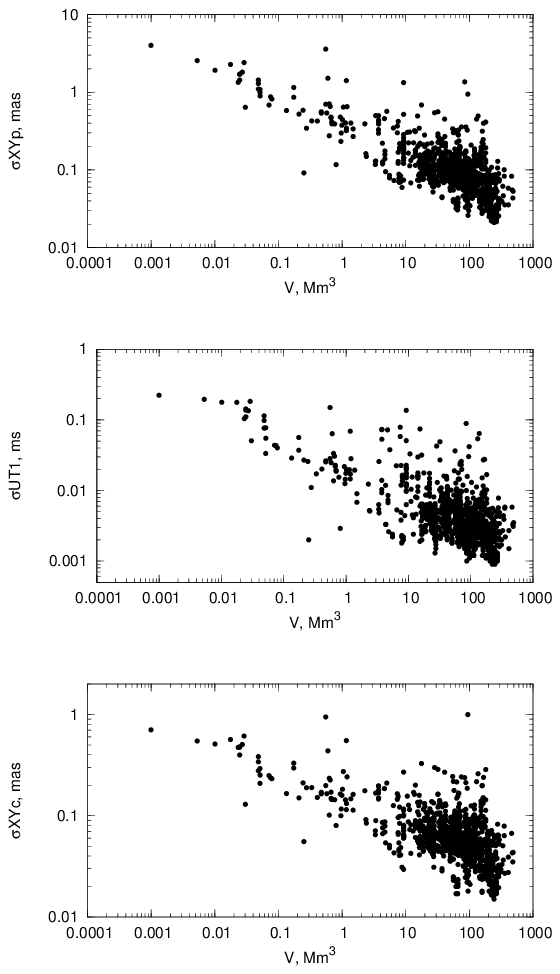}}
\caption{Dependence of the EOP precision on the network volume~$V$.
Downward, average formal error in the terrestrial pole coordinates
$X_p$ and $Y_p$, Universal Time $UT1$, and celestial pole coordinates $X_c$ and $Y_c$
are depicted (logarithmic scales on both axes)}
\label{fig:prec_all}
\end{figure}

Now, if we plot the EOP precision as a function of the network volume, as shown
in Fig.~\ref{fig:prec_all}, a clear linear (in log-log scale) dependence between
these values can be seen.

To investigate the dependence of the EOP errors on the network volume
in detail the following computations were performed.
\begin{enumerate}
\item The EOP series mentioned above was arranged in 9 groups by the network volume $V$
    in Mm$^3$:
    $V$$<$$0.1$, \ $(\sqrt{10})^k$$\le$$V$$<$$(\sqrt{10})^{k+1}$, \ $k=-2,\ldots,5$;
\item For the sessions fallen into each group the following quantities were calculated:
   \begin{itemize}
   \item the average network volume;
   \item the average formal errors in $X_p$, $Y_p$, $UT1$, $X_c$ and $Y_c$ regarded as EOP precision;
   \item weighted root-mean-square (WRMS) of the differences in $X_p$, $Y_p$ between
     the computed EOP and the IGS combined series
     igs00p03.erp\footnote{ftp://cddis.gsfc.nasa.gov/gps/products/}
     after removing the constant bias between the VLBI and IGS EOP series
     regarded as EOP accuracy.
   \end{itemize}
\end{enumerate}

\begin{table}
\centering
\caption{Results of splitting of the VLBI EOP series into 9 groups by the network volume $V$.
  For each $V$ range, number of sessions N is given, and for each parameter, minimal, maximal
  and average values are shown downward. Unit for  $X_p$, $Y_p$, $X_c$, $Y_c$: mas, unit for $UT1$: 0.1 ms}
\label{tab:bins}
\tabcolsep=4.5pt
\begin{tabular}{ccccccccp{0.2ex}cc}
\hline
V range, Mm$^3$    & N & V, Mm$^3$ & \multicolumn{5}{c}{EOP precision} && \multicolumn{2}{c}{EOP accuracy} \\
\cline{4-8}\cline{10-11}
                   &   &           & $X_p$ & $Y_p$ & $UT1$ & $X_c$ & $Y_c$ && $X_p$ & $Y_p$ \\
\hline
                     &     & 0.001 & 0.505 & 0.661 & 0.335 & 0.131 & 0.128 &&       &       \\
$V$$<0.1$            &  20 & 0.079 & 5.213 & 3.230 & 2.231 & 0.629 & 0.780 && 3.118 & 3.033 \\
                     &     & 0.037 & 1.599 & 1.500 & 1.109 & 0.383 & 0.404 &&       &       \\
\hline
                     &     & 0.134 & 0.115 & 0.068 & 0.020 & 0.056 & 0.055 &&       &       \\
0.1$\leq$$V$$<$0.316 &   7 & 0.276 & 0.889 & 1.415 & 0.564 & 0.346 & 0.315 && 0.415 & 0.385 \\
                     &     & 0.209 & 0.516 & 0.667 & 0.268 & 0.209 & 0.190 &&       &       \\
\hline
                     &     & 0.333 & 0.138 & 0.096 & 0.029 & 0.080 & 0.080 &&       &       \\
0.316$\leq$$V$$<$1   &  21 & 0.992 & 3.451 & 3.735 & 1.495 & 0.910 & 0.979 && 0.726 & 0.491 \\
                     &     & 0.667 & 0.608 & 0.686 & 0.286 & 0.208 & 0.207 &&       &       \\
\hline
                     &     & 1.01  & 0.149 & 0.119 & 0.051 & 0.084 & 0.085 &&       &       \\
1$\leq$$V$$<$3.16    &  13 & 2.40  & 1.229 & 1.599 & 0.691 & 0.617 & 0.487 && 0.481 & 0.482 \\
                     &     & 1.48  & 0.431 & 0.482 & 0.189 & 0.191 & 0.180 &&       &       \\
\hline
                     &     & 3.33  & 0.052 & 0.067 & 0.018 & 0.031 & 0.025 &&       &       \\
3.16$\leq$$V$$<$10   &  59 & 9.24  & 1.380 & 1.282 & 1.363 & 0.267 & 0.272 && 0.251 & 0.237 \\
                     &     & 6.98  & 0.232 & 0.228 & 0.195 & 0.095 & 0.094 &&       &       \\
\hline
                     &     & 10.6  & 0.052 & 0.047 & 0.013 & 0.030 & 0.032 &&       &       \\
10$\leq$$V$$<$31.6   & 188 & 31.2  & 0.835 & 0.627 & 0.744 & 0.303 & 0.353 && 0.221 & 0.188 \\
                     &     & 22.1  & 0.136 & 0.120 & 0.059 & 0.075 & 0.076 &&       &       \\
\hline
                     &     & 32.0  & 0.026 & 0.026 & 0.010 & 0.017 & 0.017 &&       &       \\
31.6$\leq$$V$$<$100  & 698 & 99.9  & 2.355 & 0.905 & 0.890 & 1.000 & 0.995 && 0.186 & 0.173 \\
                     &     & 65.4  & 0.114 & 0.094 & 0.049 & 0.065 & 0.065 &&       &       \\
\hline
                     &     & 101   & 0.021 & 0.020 & 0.009 & 0.015 & 0.015 &&       &       \\
100$\leq$$V$$<$316   & 422 & 313   & 0.678 & 0.402 & 0.642 & 0.285 & 0.287 && 0.116 & 0.129 \\
                     &     & 170   & 0.072 & 0.069 & 0.035 & 0.049 & 0.049 &&       &       \\
\hline
                     &     & 325   & 0.030 & 0.034 & 0.015 & 0.027 & 0.027 &&       &       \\
316$\leq$$V$$<$1000  &  12 & 487   & 0.091 & 0.110 & 0.058 & 0.079 & 0.079 && 0.106 & 0.080 \\
                     &     & 399   & 0.055 & 0.061 & 0.037 & 0.047 & 0.043 &&       &       \\
\hline
\end{tabular}
\end{table}

\begin{figure}
\centering
\resizebox{0.45\hsize}{!}{\includegraphics[clip]{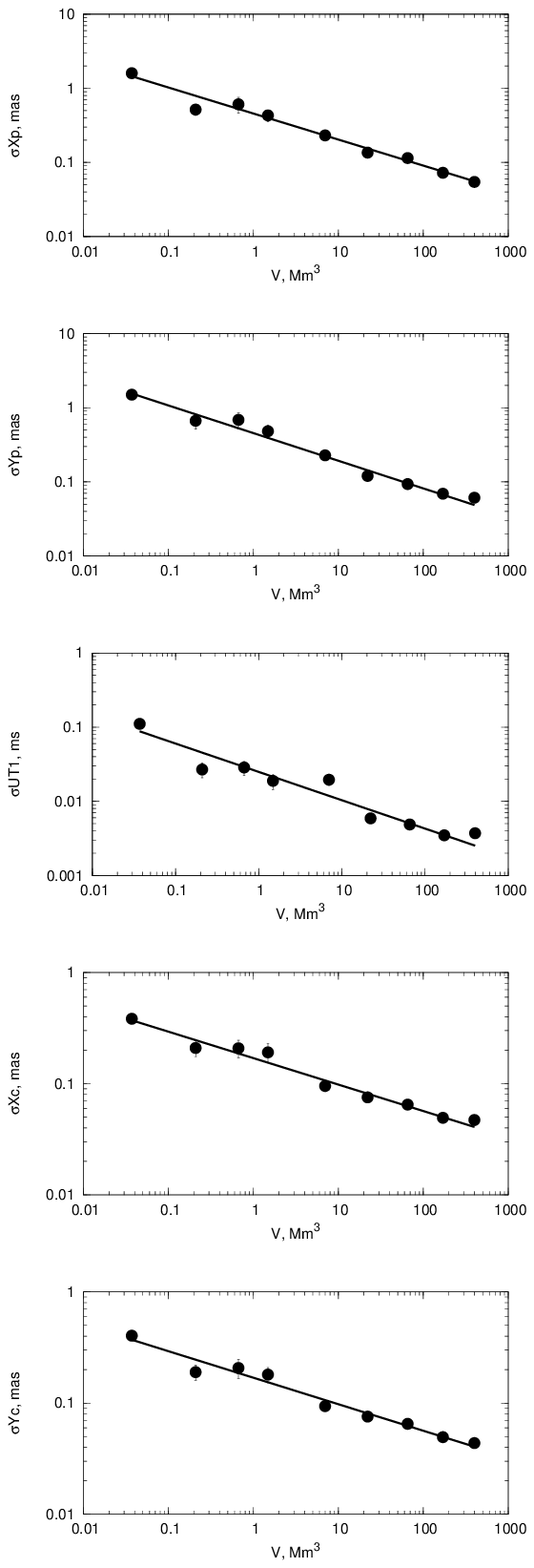}}
\caption{Dependence of the EOP precision (mean formal errors for the 9 groups of volume)
 on the network volume~$V$ (logarithmic scales on both axes).
 Solid line corresponds to the power law (Table~\ref{tab:uncerertainty_vs_v_bin})}
\label{fig:uncertainty_bin}
\end{figure}

\begin{figure}
\centering
\resizebox{0.7\hsize}{!}{\includegraphics[clip]{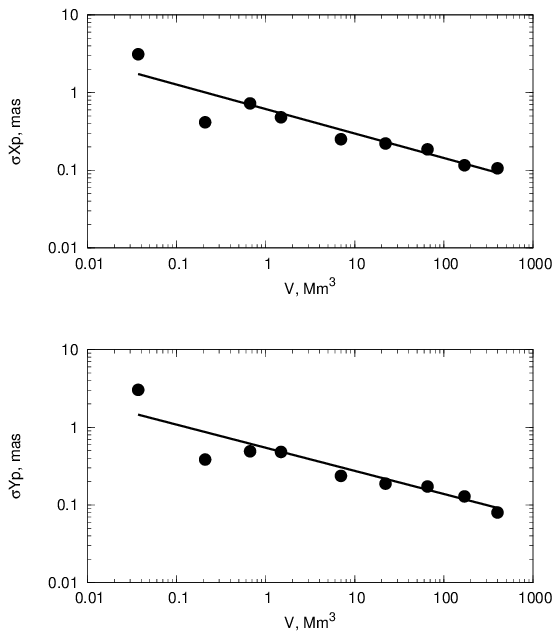}}
\caption{Dependence of the EOP accuracy (WRMS of the differences with respect to IGS
 EOP for the 9 groups of volume) on the network volume~$V$ (logarithmic scales on both axes).
 Solid line corresponds to the power law (Table~\ref{tab:accuracy_vs_v_bin})}
\label{fig:vlbi-igs_bin}
\end{figure}

Details and results of computation are shown in Table~\ref{tab:bins}
and Figs.~\ref{fig:uncertainty_bin} and~\ref{fig:vlbi-igs_bin}.
The nine averaged points thus obtained were then used to compute the parameters of a power
law fitting best to the data
\begin{equation}
\sigma=aV^c\,,
\label{eq:sigma}
\end{equation}
where $\sigma$ is an EOP error, i.e. can stand for precision or accuracy.
For actual computation this dependence was transformed to the linear form
\begin{equation}
\log\sigma = b + c\,\log V \,,
\label{eq:sigma_log}
\end{equation}
where $b=\log a$.
The parameters $b$ and $c$ were computed by means of the least square linear fit
with weights dependent on number of sessions fallen into each group.
The results are presented in
Tables~\ref{tab:uncerertainty_vs_v_bin} and~\ref{tab:accuracy_vs_v_bin},
and depicted as the regression lines in Figs.~\ref{fig:uncertainty_bin} and \ref{fig:vlbi-igs_bin}.
One can see that the dependence of both EOP precision and accuracy on the network volume
can be nicely described by a power law.

\begin{table}
\centering
\caption{Representation of the EOP precision by a power law $\log\sigma = b + c\,\log V$}
\label{tab:uncerertainty_vs_v_bin}
\begin{tabular}{cccc}
\hline
EOP & Unit & $b$ & $c$ \\
\hline
$X_p$ & mas & $-0.340  \pm 0.022$  & $-0.351  \pm 0.018$  \\
$Y_p$ & mas & $-0.342  \pm 0.023$  & $-0.373  \pm 0.019$  \\
$UT1$ & ms  & $-0.0603 \pm 0.0043$ & $-0.0382 \pm 0.0036$ \\
$X_c$ & mas & $-0.771  \pm 0.016$  & $-0.238  \pm 0.013$  \\
$Y_c$ & mas & $-0.772  \pm 0.016$  & $-0.238  \pm 0.013$  \\
\hline
\end{tabular}
\end{table}

\begin{table}
\centering
\caption{Representation of the EOP accuracy by a power law $\log\sigma = b + c\,\log V$}
\label{tab:accuracy_vs_v_bin}
\begin{tabular}{cccc}
\hline
EOP & Unit & $b$ & $c$ \\
\hline
$Xp$ & mas & $-0.212 \pm 0.045$ & $-0.315 \pm 0.038$ \\
$Yp$ & mas & $-0.263 \pm 0.049$ & $-0.298 \pm 0.041$ \\
\hline
\end{tabular}
\end{table}

Generally speaking, it is natural that the EOP precision improves
with larger network size (volume) since a large network has
larger projections of the network baselines on the rotation axes.
More interesting and less evident is that the EOP accuracy follows the same
power law with practically the same exponent $c$ but with different factor $a$.


\section{Comparison of observing programs}
\label{prog_comp}

In this section, an application of found dependence of the EOP
errors on network geometry to a comparison of EOP obtained from different IVS
observing programs will be considered.

To investigate the dependence of the EOP errors on the observing program,
we have performed the same computations as described in the previous section,
with the only difference
that EOP results were arranged in 13 groups by the IVS observing programs:
11 global networks R1, R4, RD, RDV, NEOS-A, CORE-A, CORE-B, CONT02, CONT05, T2, E3,
and 2 regional networks EURO, JADE.
Although the two last programs are primarily intended for geodesy researches
in relatively small geographic regions, and are not supposed to obtain
scientifically useful EOP results, they are included in this study for
the following reasons:
\begin{itemize}
\item including the relatively small regional networks in this study provides
significantly larger range of the network volume, and thus allows us to get more
reliable estimates for the slope of the log-log regression line
corresponding to the power law discussed above;
\item some IVS analysis centers include EURO, and sometimes JD sessions
in their EOP series.
\end{itemize}
The results shown in the previous section show that taking the small
networks into consideration does not corrupt the estimates of the
parameters of the power law describing the dependence of the EOP errors
on the network volume, which can be seen in
Figs.~\ref{fig:uncertainty_bin} and \ref{fig:vlbi-igs_bin}.

A detailed description of different IVS observing programs can be found
at the IVS website\footnote{http://ivscc.gsfc.nasa.gov/program/descrip.html},
and their main relevant characteristics are shown in Table~\ref{tab:programs}.
Note that in further computation the average value of the extremes was taken
for observing programs with changing data rate.

\begin{table*}
\begin{center}
\caption{Characteristics of IVS observing programs. Only the programs and
observations used in this paper are shown. In some columns,
the average value is given in parentheses}
\label{tab:programs}
\tabcolsep=2pt
\begin{tabular}{lcccccc}
\hline
Program    &Observation&Number  &Number          & Number of            & Network volume,& Data rate, \\
          &period   &of sessions& of stations    &observations          & Mm$^3$         & Mbps       \\
\hline
R1        & 2002--2007 & 261    &  4--8 (6.2)    & 700--4800 (2000)     & 2.4--320 (140) & 256 \\
R4        & 2002--2007 & 258    &  4--8 (6.3)    & 400--3200 (1300)     & 7.7--273 (84)  & 56--128 \\
RD        & 2004--2006 & 18     &  5--7 (6.1)    & 1200--3100 (2300)    & 16--27 (22)    & 1024 \\
RDV       & 1997--2007 & 55     &  14--20 (17.5) & 7600--29300 (17500)  & 28--313 (160)  & 128 \\
NEOS-A    & 1996--2001 & 283    &  4--6 (5.1)    & 300--2000 (1100)     & 4.7--117 (54)  & 56 \\
CORE-A    & 1997--2000 & 79     &  4--6 (5.5)    & 400--2000 (1300)     & 1.0--187 (94)  & 56 \\
CORE-B    & 1997--2001 & 53     &  4--8 (6.2)    & 400--3500 (1600)     & 0.3--235 (76)  & 56 \\
CONT02    & 2002       & 15     &  7--8 (7.9)    & 2300--3500 (3000)    & 91--92 (92)    & 256 \\
CONT05    & 2005       & 15     &  10--11 (10.9) & 5600--6500 (6000)    & 263--265 (265) & 256 \\
T2        & 2002--2006 & 41     &  5--15 (9.0)   & 300--5600 (1800)     & 11--487 (219)  & 64--128 \\
E3        & 2002--2006 & 31     &  4--6 (4.5)    & 200--800 (400)       & 1.0--88 (25)   & 128 \\
EURO      & 1996--2006 & 36     &  4--9 (6.6)    & 400--5400 (2800)     & 0.01--1.4 (0.6)& 64--128 \\
JADE      & 1999--2006 & 12     &  4--6 (5.1)    & 700--2100 (1400)  & 0.001--0.05 (0.03)& 128 \\
\hline
\end{tabular}
\end{center}
\end{table*}

For all the observing programs listed above, except RD (Research \& Development experiments),
all the sessions observed during the used time span were included in this study.
As to the RD program, only 18 sessions observed with the recording data rate 1~Gbps were used.
It is worth mentioning that CONT02 and CONT05 programs were special 2-week continuous
observing campaigns primarily aimed at the highest EOP quality.
It can also be mentioned that all observing programs were observed with the Mark~IV
terminal developed in the USA (Whitney 1998), except E3 sessions observed with the S2 terminal
developed in Canada (Wietfeldt 1996), and JADE sessions observed with the K4 terminal developed
in Japan and compatible with Mark~IV (Kiuchi 1997).
An overview and comparison of VLBI terminals can be found e.g. in Petrachenko (2000).
The S2 terminal provides maximum recording rate of 128~Mbps,
whereas the other terminals use the recording rate of 256~Mbps for regular experiments
and up to 1~Gbps in RD sessions, as can be seen in Table~\ref{tab:programs}.

\begin{figure}
\centering
\resizebox{0.45\hsize}{!}{\includegraphics[clip]{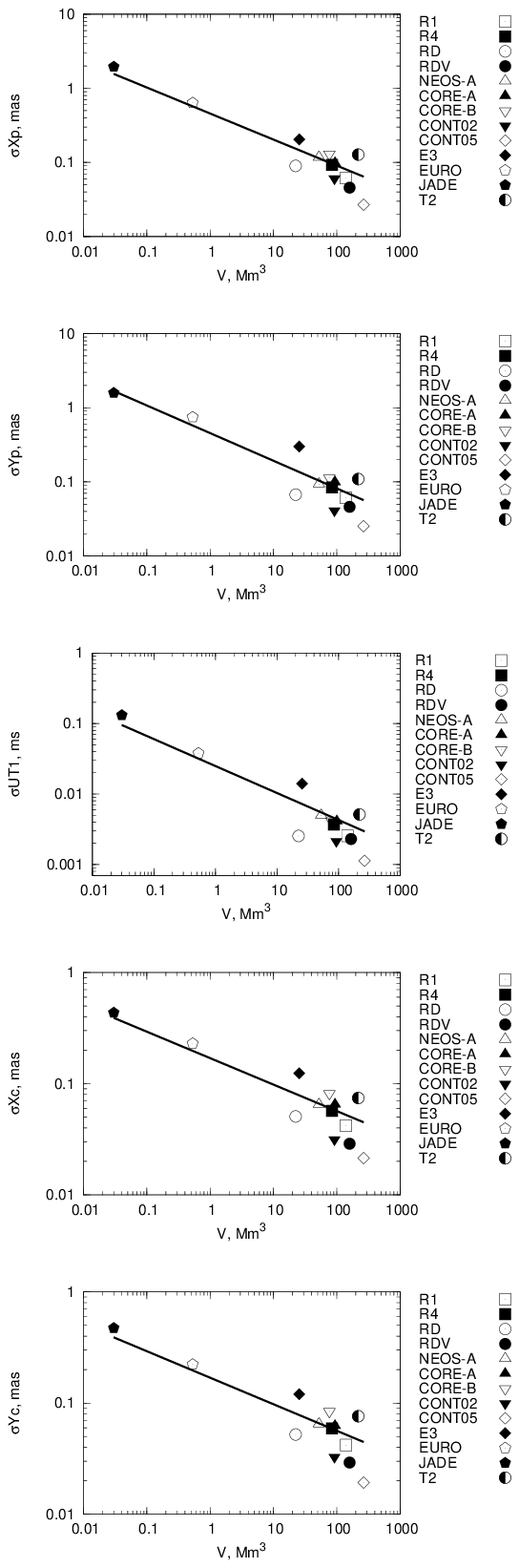}}
\caption{Dependence of the EOP precision on the average network volume~$V$
  for different observing programs (logarithmic scales on both axes).
  Solid line corresponds to the power law (Table~\ref{tab:uncerertainty_vs_v_bin})}
\label{fig:uncertainty_prog}
\end{figure}

\begin{figure}
\centering
\resizebox{0.7\hsize}{!}{\includegraphics[clip]{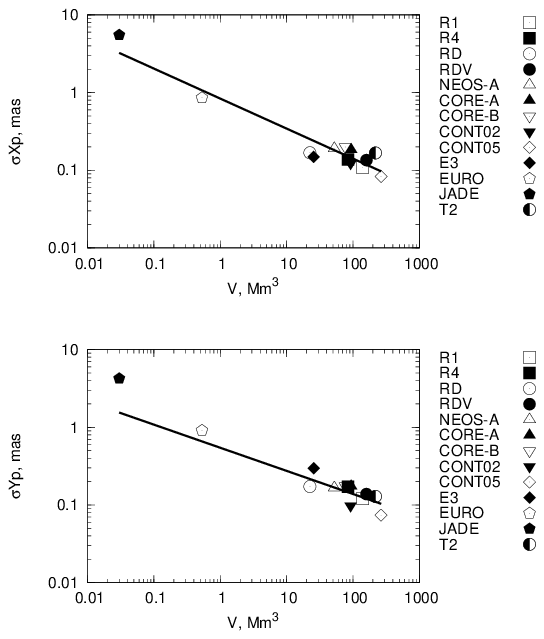}}
\caption{Dependence of the EOP accuracy on the average network volume~$V$
  for different observing programs (logarithmic scales on both axes).
  Solid line corresponds to the power law (Table~\ref{tab:accuracy_vs_v_bin})}
\label{fig:vlbi-igs_prog}
\end{figure}

The results of computations are presented in
Figs.~\ref{fig:uncertainty_prog} and~\ref{fig:vlbi-igs_prog}, where
the solid lines correspond to the power law with the parameters given in
Tables~\ref{tab:uncerertainty_vs_v_bin} and~\ref{tab:accuracy_vs_v_bin}
for EOP precision and accuracy, respectively.
Comparing these results with those presented in the previous section,
one can see that the scatter of the points with respect to the regression line
in Figs.~\ref{fig:uncertainty_prog} and~\ref{fig:vlbi-igs_prog}
is greater than that in Figs.~\ref{fig:uncertainty_bin} and~\ref{fig:vlbi-igs_bin},
which can be explained by the fact that, as a rule, networks
of different size participated in the same observing program,
which can be clearly seen from Table~\ref{tab:programs}.
Besides, the difference in some scheduling options, such as optimization,
recording mode and source selection also influence the EOP errors
and cause their deviation from the common (average) law.

Collating the results obtained above for the EOP precision and accuracy
one can see that the scatter (WRMS of the residuals) of the EOP accuracy with respect to the
regression line (0.033~mas) is much less than that of the EOP precision (0.054~mas).
The difference becomes even larger if the WRMS is scaled to the magnitude of the accuracy and
precision, respectively, more precisely to the $a$ factor in Eq.~(\ref{eq:sigma})~--- 0.039 vs. 0.108.
This may mean that the EOP accuracy follows more strongly a power law than the EOP precision.
In other words, one can conclude that the EOP accuracy
depends primarily on the network geometry and to less extent on other factors,
which have stronger impact on the EOP precision.

An analysis of the data presented in Table~\ref{tab:programs} and
Figs.~\ref{fig:uncertainty_prog} and~\ref{fig:vlbi-igs_prog}
allows us to make some interesting perceptions. Here are two examples.
\begin{itemize}
\item The known fact of smaller errors of the EOP obtained from the R1 observations
with respect to the R4 program can be mainly explained by the difference between
the average volume of the R1 and R4 networks, 140 and 84~Mm$^3$, respectively.
\item The observations made in the framework of the E3 program, using S2
terminal, involving relatively small number of stations, and delivering relatively
small number of observations, show relatively bad precision.
However, the accuracy of the E3 EOP, after accounting for the network size,
is at a level of the accuracy of other observing programs involving
more stations, using higher recording data rate, and collecting much more observations.
\end{itemize}

\begin{figure*}
\centering
\resizebox{\hsize}{!}{\includegraphics[clip]{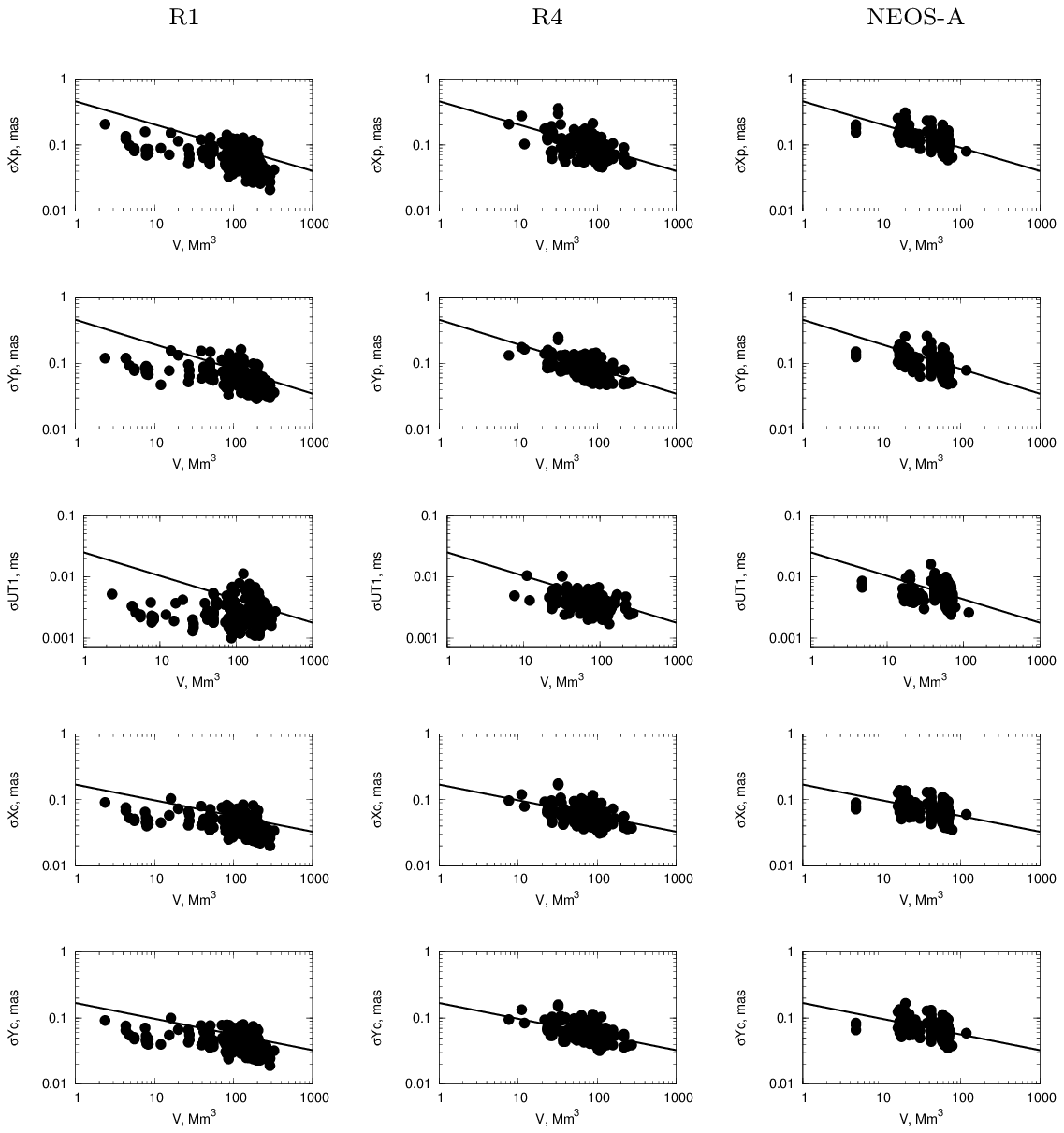}}
\caption{Dependence of the EOP precision on the average network volume~$V$
  for three IVS observing programs R1, R4 and NEOS-A (logarithmic scales on both axes).
  Solid line corresponds to the power law (Table~\ref{tab:uncerertainty_vs_v_bin})}
\label{fig:prec_R1-R4-NA}
\end{figure*}

It is interesting to see how the general law derived from the whole data set
works for separate IVS observing programs.
For this comparison, three observing programs were selected, R1, R4, and NEOS-A,
each having a large number of sessions and a large range of network volume.
All the three programs are designed for the highly accurate EOP determination
and use Mark~IV terminals, with different recording data rate however, as can be
seen from Table~\ref{tab:programs}.
Results of computations are presented in Fig.~\ref{fig:prec_R1-R4-NA}.

In these plots, one can see that the three compared observing programs
show similar slope in the dependence of the formal errors on the network
volume, and this slope is close to the general power law derived from the
whole data set.
The small discrepancies in the slope between the separate observing programs
among them and with the general slope given in Table~\ref{tab:uncerertainty_vs_v_bin}
can be evidently explained by the difference in the number of observations and
the range of the network size.
Indeed, finer dependencies of the EOP errors on the network geometry
may contribute to these discrepancies.
One can also see in Fig.~\ref{fig:prec_R1-R4-NA} that session points
for R4 and NEOS-A are laying practically on the general regression line,
whereas R1 sessions are shifted down with respect to the general law,
which corresponds to the averaged program points in Fig.~\ref{fig:uncertainty_prog}.
One of the possible reasons of a better precision of the R1 EOP may be
a higher recording data rate of 256~Mbps as compared with 56--128~Mbps
used during NEOS-A and R4 sessions.
However, possible dependence of the EOP errors on the recording data rate
is definitely a finer effect than the dependence on the network volume.

Thus, it is clear from the results presented above that to correctly
compare the errors in EOP obtained from different VLBI networks it is
necessary to properly account for the network size.
This can be done by the following method.
Before comparison of different observing programs, corresponding EOP errors
should be reduced to the unit volume.
To do so, the errors are to be modified as follows:
\begin{equation}
\sigma_m = \sigma / V^c \,,
\label{eq:mod_error}
\end{equation}
where $\sigma$ is the original error (precision or accuracy), and $\sigma_m$ is the modified error.
Original and modified errors in pole coordinates (average for $X_p$ and $Y_p$)
computed using Eq.~(\ref{eq:mod_error}) are shown in Table~\ref{tab:mod_err}.
These results give a quantitative confirmation of the preliminary conclusions
made from Figs.~\ref{fig:uncertainty_prog} and~\ref{fig:vlbi-igs_prog}.

\begin{table}
\centering
\caption{Original ($\sigma$) and modified ($\sigma_m$) errors in pole coordinates
  (average of $X_p$ and $Yp$). Unit: mas}
\label{tab:mod_err}
\begin{tabular}{lccccc}
\hline
Program & \multicolumn{2}{c}{Precision,} && \multicolumn{2}{c}{Accuracy,}    \\
        & \multicolumn{2}{c}{mas}        && \multicolumn{2}{c}{mas}\\
\cline{2-3}\cline{5-6}
          & $\sigma$ & $\sigma_m$ &&  $\sigma$ & $\sigma_m$ \\
\hline
R1        & 0.061 & 0.446 && 0.115 & 0.770 \\
R4        & 0.088 & 0.520 && 0.155 & 0.853 \\
RD        & 0.079 & 0.272 && 0.170 & 0.564 \\
RDV       & 0.046 & 0.352 && 0.137 & 0.961 \\
NEOS-A    & 0.105 & 0.512 && 0.178 & 0.816 \\
CORE-A    & 0.099 & 0.609 && 0.179 & 1.027 \\
CORE-B    & 0.121 & 0.691 && 0.188 & 0.998 \\
CONT02    & 0.051 & 0.311 && 0.111 & 0.630 \\
CONT05    & 0.026 & 0.243 && 0.079 & 0.672 \\
T2        & 0.118 & 1.021 && 0.148 & 1.178 \\
E3        & 0.253 & 0.929 && 0.223 & 0.775 \\
EURO      & 0.691 & 0.535 && 0.885 & 0.693 \\
JADE      & 1.771 & 0.435 && 4.884 & 1.267 \\
\hline
\end{tabular}
\end{table}

\begin{figure}
\centering
\resizebox{0.7\hsize}{!}{\includegraphics[clip]{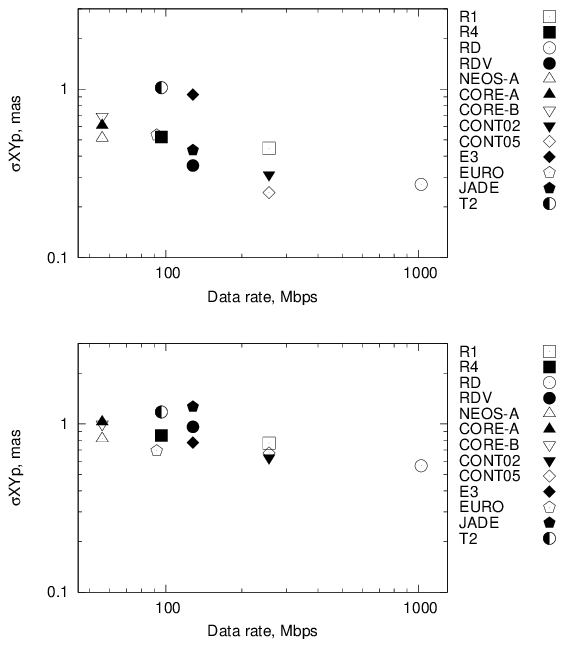}}
\caption{Dependence of the modified EOP precision ({\it top}) and accuracy ({\it bottom})
  on the recording data rate (logarithmic scales on both axes)}
\label{fig:rate_prog}
\end{figure}

Comparison of the data presented in Tables~\ref{tab:programs} and~\ref{tab:mod_err}
allows us to make a conclusion on the dependence
of the EOP errors on the recording data rate, which is depicted in Fig.~\ref{fig:rate_prog}.
From this data, one can see that the dependence of the errors in pole
coordinates on the recording data rate can also be represented by the power law:
$\log\sigma = (0.437 \pm 0.041) - (0.352 - \pm 0.121) \,\log R$
for the pole coordinates precision and
$\log\sigma = (0.338 \pm 0.023) - (0.194 - \pm 0.068) \,\log R$
for the accuracy, where $R$ is the recording data rate in Mbps.
Again, like the case of dependence of the EOP errors on the network
volume, the EOP accuracy shows stronger dependence
on the registration data rate than the EOP precision.
This can be seen from comparison of the ratio between the largest
and smallest modified values of precision and accuracy.

\section{Conclusion}
\label{concl}

The volume of a VLBI network, i.e. the volume of a polyhedron formed
by the network stations, can serve as a generalized index of network geometry.
A comparison of EOP obtained on different observing networks allowed
us to show that both EOP precision (formal errors) and accuracy
(WRMS with IGS EOP series) strongly follow
a power law with respect to the network volume in a wide range of the network size,
from regional to global one.
Thus, the network volume seems to be a governing factor
of the errors in the VLBI EOP.

Indeed, the dependence found in this paper cannot predict errors in EOP
derived from a particular VLBI session, which depend on many reasons, such as
details of the network geometry and baseline orientation, scheduling parameters
(optimization, recording mode and data rate, source selection), station operation
quality, etc., and may differ from the computed value by a factor of 2--3, as can
be seen from Fig.~\ref{fig:prec_all}.
It is intended for overall comparison of the observing networks.

Based on this result, the method is proposed to correctly compare
the EOP errors, both precision and accuracy, for different VLBI networks
and observing programs.
Following this method, the errors should be compared after a reduction
to the unit volume in accordance with Eq.~(\ref{eq:mod_error}).
After that, comparison of the modified EOP errors obtained from observations
on different networks (observing programs) does not depend on the network size anymore,
which allows us to estimate the influence of other factors on EOP results.
In particular, dependence of the EOP errors on the recording data rate was
derived from an analysis of the observations made in the framework of
several IVS observing programs.
It was found that this dependence can also be described by a power law.

It is remarkable that dependence of the EOP accuracy on both the network volume
and the recording data rate is described by a power law more accurately than
that for the EOP precision.
This may show evidence that the EOP accuracy depends mainly on the network geometry,
and the EOP precision depends to greater extent on other factors mentioned above.

The results obtained in this paper can also be used for planning geodetic
VLBI campaigns, optimal use of the existing IVS network and for designing new VLBI
networks, such as VLBI2010 (Petrachenko et al. 2004) and GGOS (Global Geodetic
Observing System, Pearlman et al. 2006, 2007).

Of course, the proposed method of accounting for the network geometry is applicable
only for the networks comprising four and more stations, but this does not
seem to be a serious limitation since the most valuable scientific VLBI
results are obtained on ramified networks.


{\bf{Acknowledgements} \ The author greatly acknowledge the hard work of the IVS components,
providing the VLBI data used in this work.
I am also very grateful to Volker Tesmer and two anonymous reviewers for
valuable critical remarks and suggestions which much helped to improve
the manuscript.



\begin{thebibliography}{}

\bibitem{Dermanis78}
Dermanis A, Mueller II (1978)
Earth rotation and network geometry optimization for very long baseline interferometers.
Bull Geod 52:131--158

\bibitem{Dow05}
Dow JM, Neilan RE, Gendt G (2005)
The International GPS Service (IGS): celebrating the 10th anniversary and looking to the next decade.
Adv Space Res 36:320--326

\bibitem{Gambis02}
Gambis D (2002)
Allan variance in Earth rotation time series analysis.
Adv Space Res 30:207-212

\bibitem{Joe91}
Joe B (1991)
GEOMPACK~--- a software package for the generation of meshes
using geometric algorithms.
Adv. Eng. Software 13:325--331

\bibitem{Johnson04}
Johnson TJ (2004)
A comparison of VLBI Earth orientation parameters from recent R1 and R4 experiments.
Geophysical Research Abstracts, 6, EGU04-A-04565

\bibitem{Kiuchi97}
Kiuchi H, Amagai J, Hama S, Imae M (1997)
K-4 VLBI data-acquisition system. Publ. Astron. Soc. Japan 49:699--708

\bibitem{Lambert06}
Lambert SB, Gontier A-M (2006)
A comparison of R1 and R4 IVS networks.
In: Behrend~D, Baver~KD (eds), Proceedings of the IVS 2006 General Meeting, Concepci\'on, Chile,
January, pp~264--268

\bibitem{Ma78}
Ma C (1978) Very long baseline interferometry applied to polar motion, relativity and geodesy.
Ph.D. Thesis, NASA TM 79582, Greenbelt, MD, USA, 367~p.

\bibitem{MacMillan00}
MacMillan D, Ma C (2000)
Improvement of VLBI EOP accuracy and precision.
In: Vandenberg~NR, Baver~KD (eds), Proceedings of the IVS 2000 General Meeting, K\"otzting, Germany,
February, pp~247--251

\bibitem{Malkin07}
Malkin Z (2007)
On dependence of EOP precision and accuracy on VLBI network.
In: Boehm J, Pany A, Schuh H (eds.), Proceedings of the  18th European VLBI for
Geodesy and Astrometry Working Meeting, Vienna, Austria, April,
Geowissenschaftliche Mitteilungen, Heft Nr. 79, Schriftenreihe der
Studienrichtung Vermessung und Geoinformation, Technische Universitaet
Wien, pp~75-78

\bibitem{Malkin08}
Malkin Z (2008)
On accuracy assessment of celestial reference frame realizations.
J Geod 82:325-329

\bibitem{Nothnagel94}
Nothnagel A, Zhihan Q, Nicolson GD, Tomasi P (1994)
Earth orientation determinations by short duration VLBI observations.
Bull Geod 68:1

\bibitem{Pearlman06}
Pearlman M, Altamimi Z, Beck N, Forsberg R, Gurtner~W, Kenyon~S,
Behrend~D, Lemoine~FG, Ma~C, Noll~CE, Pavlis~EC, Malkin~Z, Moore~AW,
Webb~FH, Neilan~RE, Ries~JC, Rothacher~M, Willis~P (2006)
Global Geodetic Observing System---considerations for the geodetic network
infrastructure.
Geomatica 60:193-204

\bibitem{Pearlman07}
Pearlman M, Altamimi Z, Beck N, Forsberg R, Gurtner~W, Kenyon~S,
Behrend~D, Lemoine~FG, Ma~C, Noll~CE, Pavlis~EC, Malkin~Z, Moore~A,
Webb~FH, Neilan~RE, Ries~JC, Rothacher~M, Willis~P (2007)
GGOS Working Group on Ground Networks and Communications.
In: Tregoning~P, Rizos~C (eds),
Dynamic Planet---Monitoring and Understanding a Dynamic Planet with Geodetic
and Oceanographic Tools, IAG Symposia, 130, pp~711--718

\bibitem{Petrachenko00}
Petrachenko WT (2000)
VLBI data and acquisition and recording systems: a summary and comparison.
In: Vandenberg~NR, Baver~KD (eds),
Proceedings of the IVS 2000 General Meeting, K\"otzting, Germany, February, pp~76--85

\bibitem{Petrachenko04}
Petrachenko B, Corey B, Himwich E, Ma C, Malkin~Z, Niell~A,
Shaffer D, Vandenberg N (2004)
VLBI 2010: Networks and observing startegies.
In: Vandenberg~NR, Baver~KD (eds),
Proceedings of the IVS 2004 General Meeting, Ottawa, February, pp~60--64

\bibitem{Schlueter07}
Schl\"uter W,  Behrend D (2007)
The International VLBI Service for Geodesy and Astrometry (IVS): current
capabilities and future prospects.
J Geod 81:379-387

\bibitem{Schuh00}
Schuh H (2000)
Geodetic Analysis Overview.
In: Vandenberg~NR, Baver~KD (eds),
Proceedings of the IVS 2000 General Meeting, K\"otzting, Germany, February, pp~219--229

\bibitem{Searle06}
Searle A (2006)
E3 network results. In: Behrend~D, Baver~KD (eds),
Proceedings of the IVS 2006 General Meeting, Concepci\'on, Chile, January, pp~330--334

\bibitem{Sokolskaya00}
Sokolskaya M, Skurikhina E (2000)
EOP determination with OCCAM and ERA packages.
In: Vandenberg~NR, Baver~KD (eds),
Proceedings of the IVS 2000 General Meeting, K\"otzting, Germany,
February, pp~309--313

\bibitem{Vennebusch07}
Vennebusch M, B\"ockmann S, Nothnagel A (2007)
The contribution of Very Long Baseline Interferometry to ITRF2005.
J Geod 81:553-564

\bibitem{Whitney98}
Whitney AR (1998)
Mark IIIA/IV/VLBA Tape Formats, Recording Modes and Compatibility.
Revision 1.12, Mark~IV Memo \#230.2, MIT Haystack Observatory, 28 August 1998

\bibitem{Wietfeldt96}
Wietfeldt RD, Baer D, Cannon WH, Feil G, Jakovina R, Leone P, Newby PS, Tan H (1996)
The S2 very long baseline interferometry tape recorder.
IEEE Transactions on Instrumentation and Measurement 45:923--929

\end{thebibliography}
\end{document}